\documentclass[aps,pra,twocolumn,showpacs,letterpaper]{revtex4}

\usepackage{graphicx}
\usepackage{bm}
\usepackage{amssymb}
\usepackage{amsmath}
\usepackage{hyperref}
\hypersetup{
    colorlinks,
    citecolor=blue,
    urlcolor=black,
    linkcolor=blue
}

\newcommand{\dd}{\operatorname{d}\!}
\newcommand{\ee}{\text{e}}
\newcommand{\ii}{\text{i}}
\newcommand{\sfrac}[2]{{\textstyle\frac{#1}{#2}}}

\renewcommand{\eqref}[1]{Eq.~(\ref{#1})}

\begin{document}

\title{Inhomogeneous Fermi mixtures at Unitarity:\\Bogoliubov-de Gennes vs. Landau-Ginzburg}

\author{J. M. Diederix}
\email{J.M.Diederix@uu.nl}
\author{K. B. Gubbels}
\author{H. T. C. Stoof}
\affiliation{
Institute for Theoretical Physics, Utrecht University,\\
Leuvenlaan 4, 3584 CE Utrecht, The Netherlands}

\date{\today}

\begin{abstract}

We present an inhomogeneous theory for the low-temperature
properties of a resonantly interacting Fermi mixture in a trap
that goes beyond the local-density approximation. We compare
the Bogoliubov-de Gennes and a Landau-Ginzburg approach and
conclude that the latter is more appropriate when dealing with
a first-order phase transition. Our approach incorporates the
state-of-the-art knowledge on the homogeneous mixture with a
population imbalance exactly and gives good agreement with the
experimental density profiles of Shin {\it et al}.\ [Nature
{\bf 451}, 689 (2008)]. We calculate the universal surface
tension due to the observed interface between the equal-density
superfluid and the partially polarized normal state of the
mixture. We find that the exotic and gapless superfluid Sarma
phase can be stabilized at this interface, even when this phase
is unstable in the bulk of the gas.

\end{abstract}


\maketitle

\section{Introduction}

The topic of imbalanced fermionic superfluidity has a long
history in condensed matter and nuclear physics and shows
presently a strong revival with the advent of ultracold
imbalanced atomic Fermi gases. It is  closely connected to
superfluid helium-3 and superconducting films in a magnetic
field, color superconductivity in neutron stars, and asymmetric
superfluidity in nuclear matter. The ultracold atom experiments
are performed in a trap to avoid contact of the atoms with
material walls that would heat up the cloud. Due to this
trapping potential the atomic cloud is never homogeneous.
However, typically the trapping frequency corresponds to a
small energy scale, so that the inhomogeneity is not very
severe. In this case, we may use the so-called local-density
approximation (LDA). It physically implies that the gas is
considered to be locally homogeneous everywhere in the trap.
The density profile of the gas is then fully determined by the
condition of chemical equilibrium, which causes the edge of the
cloud to follow an equipotential surface of the trap.

But even if the trap frequency is small, LDA may still break
down. An important example occurs when an interface is present
in the trap due to a first-order phase transition. For a
resonantly interacting Fermi mixture with a population
imbalance in its two spin states
\cite{zwierlein2006fsi,hulet2006pps}, such interfaces were
encountered in the experiments by Partridge {\it et al}.\
\cite{hulet2006pps} and by Shin {\it et al}.\
\cite{shin2008pdt} at sufficiently low temperatures. Here LDA
predicts the occurrence of a discontinuity in the density
profiles of the two spin states, which cost an infinite amount
of energy when gradient terms are taken into account.
Experimental profiles are therefore never truly discontinuous,
but are always smeared out. An important goal of this paper is
to address this effect, which amounts to solving a strongly
interacting many-body problem beyond LDA. Due to the rich
physics of the interface, we find that new phases can be
stabilized that are thermodynamically unstable in the bulk.
This exciting aspect shares similarities with the physics of
superfluid helium-3 in a confined geometry \cite{fetter1976thm}
and spin textures at the edge of a quantum Hall ferromagnet
\cite{karlhede1996teq}.

Note that the presence of an interface also can have further
consequences. Namely, in a very elongated trap Partridge {\it
et al}. observed a strong deformation of the minority cloud at
their lowest temperatures. At higher temperatures the shape of
the atomic clouds still followed the equipotential surfaces of
the trap \cite{partridge2006dtf}. The interpretation of these
results is that only for temperatures below the tricritical
point
\cite{shin2008pdt,sarma1963oiu,combescot2004ltf,parish2007ftp,gubbels2006spt,gubbels2008rgt},
the gas shows a phase separation between a balanced superfluid
in the center of the trap and a fully polarized normal shell
around this core. The superfluid core is consequently deformed
from the trap shape due to the surface tension of the interface
between the two phases
\cite{partridge2006dtf,silva2006stu,haque2007tfc}. This causes
an even more dramatic break down of LDA. Although the above
interpretation leads to a good agreement with the experiments
of Partridge {\it et al}.\ \cite{partridge2006dtf}, a fully
microscopic understanding of the value of the surface tension
required to explain the observed deformations has still not
been obtained. Presumably closely related to this issue are a
number of fundamental differences that remain with the study by
Shin {\it et al}.\ \cite{shin2008pdt}. Most importantly, the
latter observes no deformation and finds a substantially lower
critical polarization that agrees with Monte Carlo calculations
combined with LDA.

It also appears that the interfaces between the superfluid core
and the normal state are fundamentally different for the two
experiments, which might play an important role in resolving
the remaining discrepancies. In order to investigate this
interface we need to go beyond the local density approximation.
We start doing this using the Bogoliubov-de Gennes approach,
which takes all single-particle states of the complete trapping
potential into account. We show that with the correct
self-energy corrections, this approach describes both the
superfluid and normal phase rather well. However, we explain
that due to the phase transition being of first order, this
approach fails in correctly describing the interface. We
therefore put forward a different approach, based on
Landau-Ginzburg theory which is described in the second part of
this paper. With the latter approach we perform a detailed
study of the density profiles observed by Shin {\it et al.},
where our main results are summarized in
Fig.~\ref{fig:profiles}. An important feature of our theory is
that at zero temperature it incorporates the normal and
superfluid equations of state known from homogeneous Monte
Carlo simulations
\cite{giorgini2004esf,carlson2005atf,lobo2006nsp,prokofev2008fpd}.
Also crucial for the close agreement with experiment, is that
we take the energy cost of gradients in the superfluid order
parameter into account. An intriguing consequence is that we
find a stabilization of the gapless superfluid Sarma phase in
the interface due to a smoothing of the order parameter. We
return to this exciting prospect after we have discussed the
theoretical foundations on which it is based.

\section{Bogoliubov-de Gennes}

One way to describe an inhomogeneous superfluid Fermi mixture,
is by means of the so-called Bogoliubov-de Gennes equations. To
derive these we start with the BCS action
\begin{align}\label{eq:BdG_Action}
    &S[\psi^*_{\sigma},\psi_{\sigma}] = \int\!\dd\bm x\dd t \bigg\{ \frac{|\Delta(\bm x)|^2}{V_0}  \\\nonumber
    &\ +\!\sum_{\sigma}\psi^{*}_{\sigma}(\bm x, t) \left[\ii\hbar\partial_{t}+\frac{\hbar^2\nabla^2}{2m}- V^{\text{ext}}(\bm x)+\mu_{\sigma} \right]\psi_{\sigma}(\bm x, t)\\\nonumber
    &\ - \Delta^{*}(\bm x)\psi_{-}(\bm x, t)\psi_{+}(\bm x, t) -\Delta(\bm x)\psi_{+}^{*}(\bm x, t)\psi_{-}^{*}(\bm x, t)\bigg\}\;,
\end{align}
where $\psi_{\sigma}(\bm x, t)$ is the Grassmann field
associated with the fermions and $\Delta(\bm x)$ is the BCS gap
function. The external potential $V^{\text{ext}}({\bm x})$ is
to a good approximation harmonic. The measurements of
Ref.~\cite{shin2008pdt} are performed in an elongated harmonic
trap. But since they observe roughly LDA-like behavior, i.e.,
the normal-superfluid interface is small and follows the shape
of the trap, this elongation can in first instance be scaled
away and be treated as a spherical symmetric trap. The trap we
use here is thus $V^{\text{ext}}({\bf
x})=\sfrac{1}{2}m\omega^2x^2$ with $\omega$ the effective trap
frequency. The interesting physics arises when we consider an
imbalance of the fermions. To introduce an imbalance the
chemical potential of the two fermion species is different,
$\mu_{\pm} = \mu \pm h$, where $+$($-$) is the
majority(minority) species. The chemical potential difference
$2h$ then determines the polarization or imbalance. Finally,
$V_0$ is the bare attractive interaction strength between
fermions in different pseudospin states.

The action can as usual be written in a matrix form in Nambu
space. To find the poles of the propagator, we can write down
the Bogoliubov-de Gennes equations for this action. To do this
we need to expand the Grassmann fields in its energy modes,
\begin{align}
      \begin{pmatrix}
        \psi_{+}(\bm x, t) \\
        \psi^{*}_{-}(\bm x, t) \\
      \end{pmatrix}
       \propto
      \begin{pmatrix}
        u_{\bm n}(\bm x) \\
        v_{\bm n}(\bm x) \\
      \end{pmatrix}
      \ee^{-\ii E_{\bm n} t/\hbar}\;,
\end{align}
where $\bm n$ denotes the set of three quantum numbers required
to specify the eigenstates and $E_{\bm n}$ is the energy for
that single-particle state. When we use these in the equations
of motion derived from \eqref{eq:BdG_Action}, we see that we
have a solution when the Bogoliubov-de Gennes equations are
satisfied. These are differential equations of the form
\begin{align}\label{eq:BdG_BdGEquation}
      \begin{pmatrix}
        \hat{K}_{+}(\bm x) & \Delta(\bm x) \\
        \Delta^{*}(\bm x) & -\hat{K}_{-}(\bm x) \\
      \end{pmatrix}
      \cdot
      \begin{pmatrix}
        u_{\bm n}(\bm x)\\
        v_{\bm n}(\bm x) \\
      \end{pmatrix}
      =E_{\bm n}
      \begin{pmatrix}
        u_{\bm n}(\bm x)\\
        v_{\bm n}(\bm x) \\
      \end{pmatrix}\;,
\end{align}
where $\hat{K}_{\sigma} = -\frac{\hbar^2}{2m}\nabla^2+
V^{\text{ext}}-\mu_{\sigma}$. This gives a coupled set of
differential equations with the boundary conditions that both
the coherence factors $u$ and $v$ are zero at infinity and
smooth at the origin. We also have the normalization condition
$\int\dd\bm x(|u_{\bm n}|^2+|v_{\bm n}|^2)=1$ for each $\bm n$.
Only for certain discrete values of $E_{\bm n}$ these
conditions can be satisfied.

Since the trap considered by us is spherically symmetric, the
gap $\Delta$ is a function of the radius only. We therefore
write
\begin{align}\label{eq:BdG_uv_definition}
    u_{\bm n}(r,\theta,\phi) = \frac{u_{nl}(r)}{r}Y_{lm}(\theta,\phi),
\end{align}
where $Y_{lm}$ are the spherical harmonics and we do the same
for $v$. Note that the sum over $\bm n$ is now a sum over the
set $\{n,l,m\}$. With this substitution we find an equation for
the new functions $u_{nl}(r)$ and $v_{nl}(r)$. It becomes
\begin{align}\label{eq:BdG_uvDivEquation}
      \partial_r^2
      \cdot
      \begin{pmatrix}
        u_{nl}(r)\\
        v_{nl}(r) \\
      \end{pmatrix}
      =-\bm{H}_{nl}(r)\cdot
      \begin{pmatrix}
        u_{nl}(r)\\
        v_{nl}(r) \\
      \end{pmatrix}\;.
\end{align}
The matrix $\bm H$ is given by
\begin{align}\begin{split}\label{eq:BdG_Hamiltonian}
 \bm H_{nl} (r) = &\  \frac{2 m}{\hbar^2}
    \begin{pmatrix}
        \mu_{+}-V_l(r)+E_{nl}& -\Delta(r)\\
        \Delta(r)  & \!\!\!\!\!\!\!\! \mu_{-}-V_l(r)-E_{nl} \\
    \end{pmatrix},\\
    V_{l}(r) = &\ m\omega^2 \frac{r^2}{2}+\frac{\hbar^2}{2m}\frac{l(l+1)}{r^2}\;.
\end{split}\end{align}
Here $E_{nl}$ are the energies of the system and $V_l(r)$ is
the effective external potential including the effect of the
centrifugal barrier for nonzero $l$. Since the chemical
potentials are $\mu_{\pm}=\mu\pm h$, we notice that in
principle $h$ can be absorbed in the energy $E_{nl}$. When $h$
is absorbed in the energy by $E'_{nl} = E_{nl}+h$ there exists
the symmetry, $E'\rightarrow -E'$ for $(u,v)\rightarrow(-v,u)$,
which reduces the number of states we have to compute by a
factor of two, i.e., we only need the positive energy states.
The boundary conditions are that both $u_{nl}$ and $v_{nl}$
must be zero in the origin and at infinity.

The properly normalized noninteracting solutions $(u_{nlm},
v_{nlm})= (\phi_{nlm},0)$ are given by the equations
\begin{equation}\begin{split}\label{eq:3dHarmonicOscillator}
    \phi_{nlm}(r,\theta,\phi)&=\mathcal{N}(n,l)\;\ee^{-\frac{r^2}{2\ell^2}}\left(\sfrac{r}{\ell}\right)^lL_n^{l+\frac{1}{2}}(\sfrac{r^2}{\ell^2})Y_{lm}(\theta,\phi),\!\!\!\!\\
    E'_{nl}&=\hbar\omega\left(\frac{3}{2}+2n+l\right)-\mu\;,\\
    \mathcal{N}(n,l) &= \sqrt{\frac{2^{n+l+2}n!}{{\ell^3(1+2l+2n)!!\sqrt{\pi}}}}\;,
\end{split}\end{equation}
with $L_n^l$ the associated Laguerre polynomials,
$\mathcal{N}(n,l)$ a normalization constant, and
$\ell=\sqrt{\hbar/m\omega}$ the so-called trap length.

\subsection{Numerical Methods}

To solve the differential equation in
\eqref{eq:BdG_uvDivEquation} we use the so-called modified
Numerov algorithm. To explain this we first introduce the
vector $(u,v)=\bm w$ in Nambu space, and notice that, since we
are dealing with a second-order two-channel differential
equation, we can in principle find two sets of independent
solutions. We can benefit from this, because this allows us to
numerically set boundary conditions on both ends of the
solution. We will therefore solve the equation for two sets at
once and introduce for that the matrix $\bm W = (\bm w^{(1)},
\bm w^{(2)})$ with two independent sets in its columns. The
differential equation now becomes,
\begin{equation}
     \partial^2_r\cdot\bm W(r) = -\bm H(r)\cdot\bm W(r)\;.
\end{equation}
We can now discretize $\bm W$ with step size $h$ and use the
modified Numerov algorithm. This algorithm gives very accurate
solutions since the error we make is only of order $h^4$. The
recurrence relation for $\bm W$ is,
\begin{equation}\begin{split}\label{eq:BdG_Numerov}
    \bm W_n =& \left[1 + \frac{h^2}{12} \bm
    H_{n}\right]^{-1}\!\!\!\cdot\left[-\left(1 + \frac{h^2}{12}\bm
    H_{n-2}\right)\cdot\bm W_{n-2} \right. \\
    &\quad+2\left.\left(1-5\frac{h^2}{12}\bm H_{n-1}\right)\cdot\bm
    W_{n-1}\right]+\mathcal{O}(h^4)\;.
\end{split}\end{equation}
Notice that if $\bm W$ is a solution so is $\bm W\cdot\bm A$,
with $\bm A$ some arbitrary matrix. We can use this to keep the
numerical errors under control. Since both channels are closed
we analytically have an exponentially decaying but also an
exponentially growing solution. Small numerical errors tend to
trigger this growing solution and therefore give dramatically
wrong solutions. We can fix this by diagonalizing $\bm W$,
during the walk over $n$ in \eqref{eq:BdG_Numerov}, whenever an
element of $\bm W$ grows larger than a certain value, for which
we typically use 10.

The boundary conditions for the discretized solutions are
simply that they are to be zero at $r=0$ and zero at
$r=\infty$. But since the wavefunction is exponentially
suppressed (closed) in the classically forbidden region,
$r_{\infty}$ can be taken close to the classical edge. However,
for nonzero angular momentum $l$, $u$ and $v$ are closed at
both ends, making it hard to obey both boundary conditions. The
easiest way to handle this is to start in both ends with the
proper boundary conditions (the functions being zero), and glue
them together at a certain point. To find the independent
solutions we use the following boundary conditions on $\bm W$,
\begin{equation}
    \bm W_0 =
    \begin{pmatrix}
        0 & 0\\
        0 & 0\\
    \end{pmatrix}
    \qquad
    \bm W_1 =
    \begin{pmatrix}
        h & 0\\
        0 & h\\
    \end{pmatrix}\;,
\end{equation}
where $n=0$ corresponds to either $r=0$ or $r=r_{\infty}$ and
$n=1$ to $r=h$ or $r=r_{\infty}-h$ respectively. We have enough
freedom to match $v$ smoothly and only match the value of $u$
(or vice versa). Only at certain values for the energy $E_{nl}$
also the derivative of $u$ ($v$) can be matched. It is crucial
to pick a proper matching point $r_\text{m}$, in order to make
a quickly converging loop to find all the energies. A good
choice is on top of the outer maximum of $u$. A good
approximation of this maximum can be found analytically by
linearizing the potential near the classical turning point.

The equations described so far can be used to find the particle
distribution or densities given a certain gap. To find the
right gap function we need the gap equation. This equation
follows in mean-field theory from the definition of the gap,
\begin{align}\label{eq:gap_Equation}
\begin{split}
    \frac{\Delta(\bm x)}{V_0} = & \langle \psi_{-}(\bm x,t)\psi_{+}(\bm x,t) \rangle \\
    = & \sum_{\bm n} u_{\bm n}^{\phantom{*}} (\bm x)v_{\bm n}^*(\bm x) N_{\text{F}}(E_{\bm n})\;,
\end{split}
\end{align}
where $u$ and $v$ are calculated with the Bogoliubov-de Gennes
equations using the same $\Delta$ and the sum is also over the
negative energy states. The Fermi distribution is denoted by
$N_{\text{F}}(E)$. Under the symmetry of
\eqref{eq:BdG_Hamiltonian} the gap equation can be written in
the more familiar form
\begin{align}
\begin{split}
    \frac{\Delta(\bm x)}{V_0} = & - \!\!\!\!\!\sum_{\bm n|E'_{\bm n}>0}\!\!\! u_{\bm n}^{\phantom{*}}(\bm x)v_{\bm n}^*(\bm x)\ \times \\
    &\ \ \times\left[1-N_{\text{F}}(E'_{\bm n}+h)-N_{\text{F}}(E'_{\bm n}-h)\right].
\end{split}
\end{align}
The inverse of the (bare) interaction strength $V_0$ can be
written in terms of the $T$-matrix by
\begin{align}\label{eq:TMatrix}
    \frac{1}{V_0} = \frac{m}{4\hbar^2\pi a} - \frac{1}{V} \sum_{\bm k}\frac{1}{2\epsilon_{\bm k}} \;,
\end{align}
where $a$ is the scattering length and $\epsilon_{\bm
k}=\hbar^2\bm k^2/2m$ is the energy of a free particle with
momentum $\bm k$. The sum in the above equation is divergent,
however, the right-hand side of \eqref{eq:gap_Equation}
contains the same divergence for the negative energy states.
These divergences cancel to get a regular gap equation.

There is a common procedure \cite{grasso2003hfb} to handle
these divergences and in the process also improve the numerical
convergence of the gap equation. To handle the divergence, the
idea is to notice that for large negative energy $E_{\bm n}$ in
\eqref{eq:gap_Equation}, the sum can be approximated by the
integral as
\begin{align}\label{eq:tail_approximation}
   \sum_{\bm n = {\bm n}_{\text{C}}}^{\infty} u_{\bm n}^{\phantom{*}}(\bm x) v_{\bm n}^*(\bm x) \approx -\int_{k_{\text C}(\bm x)}^{\infty}\frac{\dd \bm k}{(2\pi)^3}\frac{\Delta(\bm x)}{2(\epsilon_{\bm k} - \mu(\bm x))}\;,
\end{align}
where $k_{\text C}(\bm x) = \sqrt{2m(|E_{\text C}|-V(\bm
x))/\hbar^2}$ with $E_{\text C}$ the (negative) energy
belonging to state ${\bm n}_{\text C}$ and $\mu({\bm
x})=\mu-V(\bm x)$. The difference of this integral and the one
in the left-hand side of \eqref{eq:gap_Equation} after
substituting \eqref{eq:TMatrix} is finite and can be computed
as
\begin{align}\begin{split}
    &G_{\text{Reg}}(\bm x)=\\&\quad\frac{m}{4\pi^2\hbar^2}\left(2k_{\text C}(\bm x) - k_{\text F}(\bm x) \log\frac{k_{\text C}(\bm x)+k_{\text F}(\bm x)}{k_{\text C}(\bm x)-k_{\text F}(\bm x)} \right),
\end{split}\end{align}
where $k_{\text F}(\bm x) = \sqrt{2 m \mu(\bm x)/\hbar^2}$.
This result is thus also a function of $\bm x$. The complete
gap equation is now
\begin{align}
    \Delta(\bm x) \approx \sum_{\bm n = 0}^{{\bm n}_{\text{C}}}\frac{ u_{\bm n}^{\phantom{*}}(\bm x) v_{\bm n}^*(\bm x) N_{\text{F}}(E_{\bm n})}{\frac{m}{4\pi \hbar^2 a} - G_{\text{Reg}}(\bm x)}\;,
\end{align}
where $|E_{\text{C}}| \gg \mu$ to ensure a good numerical
convergence. Fortunately, in practice a factor of the order of
ten between the energy and the chemical potential can be enough
to have reasonable convergence.

In the unitarity limit the scattering length goes to infinity,
which is a well defined limit in this description of the gap
equation. The convergence behavior of the gap equation depends
mostly on the size of $\Delta$. The individual superfluid
eigenstates differ the most from the normal eigenstates around
the Fermi level. The gap sets the energy scale for the distance
from the Fermi level where states are still affected. Since the
normal eigenstates give no contribution to
\eqref{eq:tail_approximation}, this has an immediate effect on
the convergence, i.e., the appropriate value for
$E_{\text{C}}$.

Using the eigenstates found with the Bogoliubov-de Gennes
equations, we can compute the densities directly. They are
given by
\begin{align}\begin{split}\label{eq:densities}
    n_{+}(\bm x)=&\sum_{\bm n}|u_{\bm n}(\bm x)|^2 N_{\text F}(E_{\bm n})\\
    n_{-}(\bm x)=&\sum_{\bm n}|v_{\bm n}(\bm x)|^2 N_{\text F}(-E_{\bm n})\;.
\end{split}\end{align}
The energy symmetry of \eqref{eq:BdG_Hamiltonian} reduces these
to the more familiar form
\begin{align}\begin{split}
    n_{\pm}(\bm x) = \sum_{\bm n|E'_{\bm n}>0}\ &|v_{\bm n}(\bm x)|^2\big[1-N_{\text F}(E'_{\bm n} \mp h)\big]\\
      + &|u_{\bm n}(\bm x)|^2 N_{\text F}(E'_{\bm n} \pm h)\;,
\end{split}\end{align}
where $h$ in the Bogoliubov-de Gennes equations
\eqref{eq:BdG_BdGEquation} is absorbed in the energy. This
expression does not converge as quickly as the gap in terms of
a reasonable cutoff ${\bm n}_{\text C}$. The gap equation only
converges so quickly, because of the proper approximation of
the large energy tail. But we can do the same thing for the
density expression. For low temperatures we can approximate the
tail of the sum over $|v|^2$ as
\begin{align}
    \sum_{\bm n = \bm n_{\text C}}^{\infty} |v_{\bm n}(\bm x)|^2 \approx \int_{ k_{\text C}(\bm x)}^{\infty}
    \frac{\dd \bm k}{(2\pi)^3}\frac{|\Delta(\bm x)|^2}{2(\epsilon_{\bm k} - \mu(\bm x))^2}\;,
\end{align}
where the cutoff is of course the same as for the gap equation.
This concludes the method of solving the Bogoliubov-de Gennes
equations. There is one issue remaining, namely the addition of
self-energy effects, which is very important in the strongly
interacting unitarity limit. We discuss this issue next.

\subsection{Normal phase}

The Bogoliubov-de Gennes method so far describes a superfluid
for all values of the scattering length, even an infinite one,
i.e., the unitarity limit. However, with strong interactions
there is a lot more going on than just forming Cooper pairs. In
order to take all (mean-field) interaction effects into account
we add a diagonal self-energy to the action in
\eqref{eq:BdG_Action}. Because of the very strong interaction,
it is hard to find a rigorous microscopic derivation of the
self-energy. Instead we will use the knowledge gained by the
Monte-Carlo simulations
\cite{giorgini2004esf,carlson2005atf,lobo2006nsp} and use an
effective self-energy that accurately describes these
simulations.

\begin{figure}[t]
    \includegraphics[width=\columnwidth]{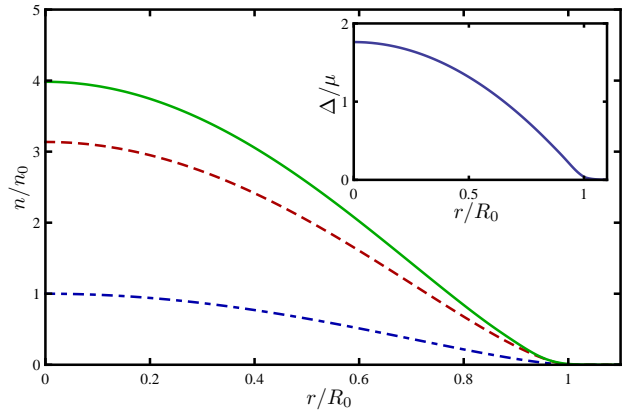}
    \caption{
    (Color online) The density profiles of the normal phase and
    the superfluid phase of a balanced unitary gas at zero temperature. Both calculated using the Bogoliubov-de
    Gennes method. Besides a small smoothening near the edge, this
    gives the same result as LDA. The dash-dotted line is the ideal-gas
    result, here $R_0$ is the radius of the ideal gas cloud and $n_0$ the
    central atomic density. The dashed line is the normal phase
    with self-energy effects and the solid line the superfluid phase.
    The inset shows the gap parameter for the superfluid phase.
    }\label{fig:BdG_Balanced}
\end{figure}

The self-energy depends on the pseudospin of the particle and
can be added to the Hamiltonian of \eqref{eq:BdG_Hamiltonian}
in the following way,
\begin{align}
    \bm H_{nl}^{\text{tot}} = \bm H_{nl} -\frac{2m}{\hbar^2} \begin{pmatrix}
                                              \hbar\Sigma_{+}(\bm x) & 0 \\
                                              0 & \hbar\Sigma_{-}(\bm x) \\
                                           \end{pmatrix}\;.
\end{align}
In the polarized case, these self-energies are really different
from each other, and cannot be written in the same form as the
chemical potential, such that the difference is a constant
independent of position. This means that the $E'\rightarrow-E'$
symmetry of the Bogoliubov-de Gennes equation is gone. The BCS
coherence functions $u_{\bm n}$ and $v_{\bm n}$ thus have to be
computed independently for positive and negative energies. For
the gap equation we can only use \eqref{eq:gap_Equation} and
for the densities only \eqref{eq:densities}.

The self-energy originates from the interaction, which is only
nonzero for particles of different species. Therefore we take
the majority(minority) self-energy proportional to the
majority(minority) atomic density. From the Monte-Carlo results
\cite{lobo2006nsp,pilati2008psp} the equation of state is known
for a highly polarized mixture at unitarity. From this equation
of state we can extract an accurate approximation to the
self-energy. When we assume that all interaction effects are
incorporated by the self-energies, the equation of state
becomes
\begin{align}\label{eq:EOS_density}
    \frac{E}{V} = \frac{3}{5} \frac{\hbar^2}{2m}(6\pi^2)^{2/3}(n_{+}^{5/3}+n_{-}^{5/3})+n_{+}n_{-}\Gamma\;,
\end{align}
where we already used that
$\hbar\Sigma_{\sigma}=n_{-\sigma}\Gamma$. Symmetry arguments
and dimensional analysis suggest the following form for the
effective interaction,
\begin{align}\label{eq:selfenergy_density}
    \Gamma  = - \frac{3}{5} \frac{\hbar^2}{2m}(6\pi^2)^{2/3}\frac{A}{(n_{+}^{\alpha} +n_{-}^{\alpha})^{1/3{\alpha}}}\;,
\end{align}
with $\alpha$ a fit parameter and $A$ can in principle be
obtained from a simple ladder calculation. This equation of
state overlaps very accurately with the Monte-Carlo equation of
state for $\alpha=3$ and $A=0.96$. The resulting self-energies
can be directly added to the Bogoliubov-de Gennes equations. In
the normal state and in the limit of large $\mu\gg\hbar\omega$,
the results of the Bogoliubov-de Gennes equations are equal to
the local density approximation, with the exception of boundary
effects as shown in Fig.~\ref{fig:BdG_Balanced}.

\subsection{Superfluid phase}

The self-energies we use for the normal phase are not correct
for the superfluid phase. In order to improve on that we can
add a superfluid correction. The simplest way to do that is to
include a second-order correction in $\Delta$. From other
work\cite{gubbels2006spt,astrakharchik2004esf,carlson2005atf}
it is known that the balanced unitary superfluid behaves like a
BCS superconductor with a scaling factor of the chemical
potential. In BCS theory, the relation between the chemical
potential and the Fermi energy is $\mu =
(1+\beta_{\text{BCS}})\epsilon_{\text F}$ with
$\beta_{\text{BCS}}=-0.41$ and $\epsilon_{\text
F}=\hbar^2(6\pi^2 n)^{2/3}/2m$. For the unitarity Fermi gas
this relation is $\mu=(1+\beta)\epsilon_{\text F}$ with
$\beta=-0.59$. This scaling can be incorporated in the
self-energy as $\mu' = \mu - \hbar\Sigma$, where $\mu'$ then
plays the role of the chemical potential in BCS theory, from
which it follows that
$\mu'=(1+\beta_{\text{BCS}})\epsilon_{\text F}$. We thus obtain
that $\hbar\Sigma = (\beta-\beta_{\text{BCS}})\epsilon_{\text
F}$. The self-energy we find using this scaling, can be
compared with the equal density one from
\eqref{eq:selfenergy_density}. It follows that both
self-energies are equal for $A_{\text{sf}}=0.32$. The $\alpha$
is not fixed with the scaling approach and therefore we use
$\alpha=3$. This smaller value of $A$ shows that the
\emph{normal} self-energy overestimates the diagonal
self-energy in the superfluid phase.

In this paper we want to investigate what happens at the
surface between the normal and the superfluid phase in the
imbalanced trapped Fermi mixture. To do this we need the
self-energy not only in the equilibrium normal and superfluid
phase separately, but also out of equilibrium. We achieve this
by considering a $|\Delta|^2$ correction to the self-energy. We
thus write for the effective interaction,
\begin{align}\label{eq:BdG_SF_SelfEnergy}
    \Gamma  = - \frac{3}{5} \frac{\hbar^2}{2m}(6\pi^2)^{2/3}\frac{A - (A - A_{\text{sf}}) \frac{|\Delta|^2}{|\Delta_0|^2}}{(n_{+}^3 +n_{-}^3)^{1/9}}\;,
\end{align}
here $|\Delta_0| = 1.67\mu$ is the value of the gap in the
balanced superfluid for our modified BCS theory with
self-energy effects. Within Bogoliubov-de Gennes theory this
self-energy now incorporates the correct equation of state for
both the normal and the superfluid phase. With this we can thus
study the behavior around the interface.

\begin{figure}[t]
    \includegraphics[width=\columnwidth]{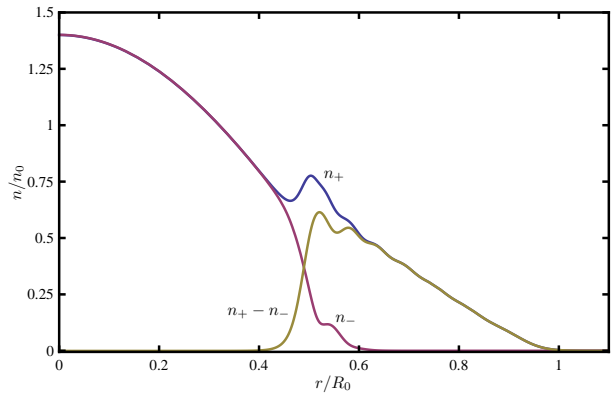}
    \caption{
    (Color online) The density profiles of the polarized superfluid in the unitarity limit at zero temperature
    calculated using the Bogoliubov-de Gennes method. The interface between the normal
    and superfluid phase is clearly visible, and has a nonzero width.
    The minority density, however, drops quickly to zero after the interface,
    which is in contrast with experiments. We used $\mu=21\hbar\omega$ and about $P=0.48$ for
    the polarization. The scaling is as in Fig.~\ref{fig:BdG_Balanced}, with $R_0$ and $n_0$ for
    the majority species.
    }\label{fig:BdG_Polarized}
\end{figure}

\begin{figure}[t]
    \includegraphics[width=\columnwidth]{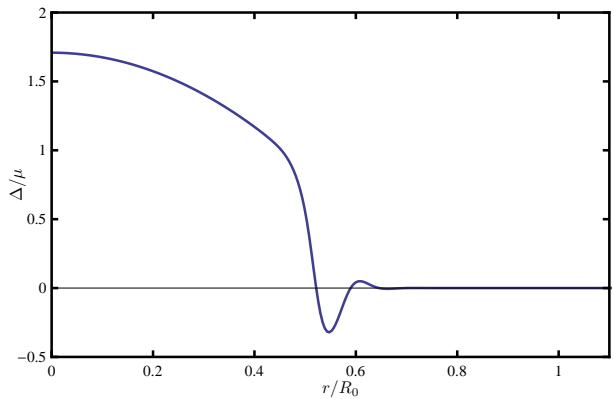}
    \caption{
    (Color online) The gap parameter for a polarized superfluid in the unitarity limit
    calculated using the Bogoliubov-de Gennes method. The
    interface between the normal and superfluid phase is clearly visible,
    and has a nonzero width. Near the interface, oscillations are clearly
    visible. The physical parameters are the same as in Fig.~\ref{fig:BdG_Polarized}.
    }\label{fig:BdG_Polarized_gap}
\end{figure}

An interface appears when we consider a population imbalance in
the Fermi mixture. We can arrive at this by setting the
chemical potential difference $h$ to a nonzero value. In
principle the Bogoliubov-de Gennes method solves such a system,
however, in practice there are technical details that decrease
the efficiency of the iterative process used to find a solution
to the gap equation. The major difficulty is that different
energy levels can get close together, such that they are not
easily distinguished. In order to find all energy levels up to
the cutoff, a very good guess of the energy is needed to start
with. To get a good guess, we start with the gap and
self-energy set to zero, so that the energies are given by
\eqref{eq:3dHarmonicOscillator}, and slowly increase them. This
method, although time consuming, works very well. However, this
in turn gives rise to avoided crossings that complicates the
algorithm to guess the energies. The resulting algorithm is
rather slow, but fast enough to give results after a few hours
of computer running time.

\subsection{Results}

The approach using the Bogoliubov-de Gennes equations together
with the gap equation, works well in the sense that it
converges to a reasonable function for the gap. Most of the
results plotted in this paper are performed with the chemical
potential $\mu=21\hbar\omega$. This value is chosen for
numerical convenience, since it is large enough that most
finite-size effects are small, but not too large such that a
result can be found within a reasonable amount of time. This
value of the chemical potential corresponds to about
$2\times10^4$ particles, which is somewhat less than used in
experiments that have about $10^5$-$10^7$. The dependence on
the total number of atoms can, however, be scaled away in the
unitarity limit, as is done in the figures as well as with the
data of MIT. Only finite-size effects near the edge or the
interface are affected by the total number of atoms.

\begin{figure}[t]
    \includegraphics[width=\columnwidth]{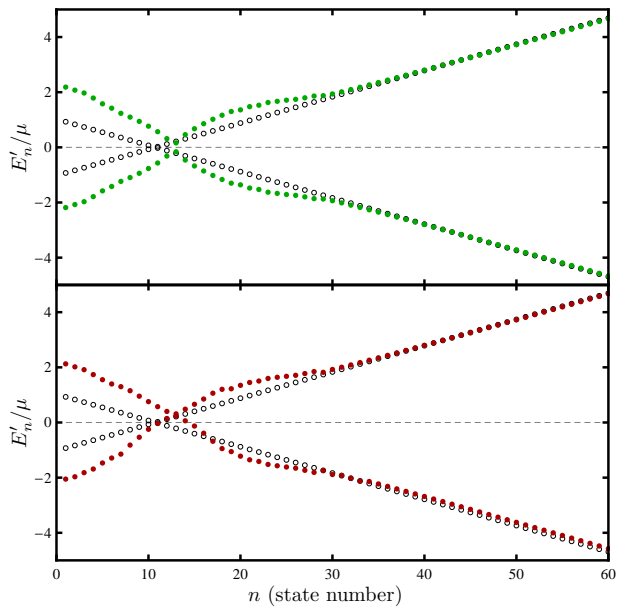}
    \caption{
    (Color online) The energy spectrum in the unitarity limit for zero angular momentum $l=0$
    for the normal state (open dots), the balanced superfluid (upper figure) and
    the polarized superfluid (lower figure).
    }\label{fig:BdG_Energy_spectrum}
\end{figure}

\begin{figure}[t]
    \includegraphics[width=\columnwidth]{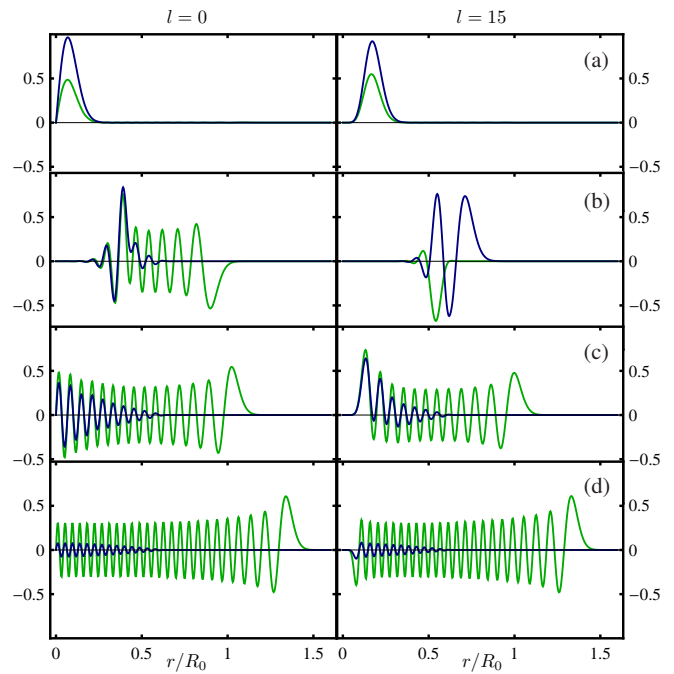}
    \caption{
    (Color online) Some examples of eigenstates of the trapped
    Fermi mixture. The two lines are the $u_{nl}(r)$ (light green line) and $v_{nl}(r)$ (dark blue line) functions
    from the Bogoliubov transformation in \eqref{eq:BdG_uv_definition}. In the left column
    states with angular momentum zero $l=0$, and in the right column with $l=15$.
    On the rows (a) through (d), states with increasing $n$ are shown.
    The state in (b) is near the Fermi-surface and thus deviates most from the
    harmonic oscillator states, while in (d) the $n$ is very large and the states look
    more like the harmonic oscillator states.
    }\label{fig:BdG_Energy_states}
\end{figure}

In Fig.~\ref{fig:BdG_Energy_states} the $u$'s and $v$'s for
some one-particle eigenstates are plotted. This shows that the
states near the Fermi surface deviate substantially from the
harmonic oscillator states, whereas for larger energies they
become more alike. The fact that the harmonic oscillator states
are a good approximation for large energies is used when
dealing with the regularization of the divergence in the gap
equation \eqref{eq:tail_approximation}. In Fig.
\ref{fig:BdG_Energy_spectrum} the energy spectrum for $l=0$ is
shown for both the polarized and the balanced superfluid. The
polarized energies are clearly not symmetric for $E'_n
\rightarrow\ -E'_{-n}$ whereas the balanced energies have this
symmetry. Notice that in both cases, the chemical potential
difference $h$ is absorbed in the energies. The difference
between the \emph{positive} en \emph{negative} branch is thus
entirely caused by the difference in the self-energies.

The Bogoliubov-de Gennes equations with the gap equation and
the self-energy effects included, seems to describe the
observed physics rather well. However, a closer look shows
problems with this method. These problems are critical when
looking at the interface in the Fermi mixture. The two most
important problems are the oscillations that appear near the
interface and the incorrect location of the interface itself.

The location of the interface is determined by the gap
equation, which in normal BCS theory minimizes the
thermodynamic potential. However, in the polarized case, the
phase transition is a first-order transition. This means that
the thermodynamic potential close to the phase transition has
two minima and a maximum, all of which are a solution to the
gap equation. The real transition should occur when the value
of the thermodynamic potential in the $\Delta=0$ minimum
becomes lower than its value in the other superfluid minimum.
This condition crucially depends on the actual value of the
thermodynamic potential in the superfluid minimum. Although
this value is known from other analysis, see
\eqref{eq:SF_energyfunc_minimum} below, it is not included in
this model. Because of this problem, the interface is shifted
outwards and as a result almost completely removes the
partially-polarized shell from the theory.

The oscillations in the gap parameter and the density profiles
are related to the proximity effect near the
interface\cite{mcmillan1968tsn}. This seems to be a common
feature in many Bogoliubov-de Gennes analysis
\cite{kinnunen2006sif,ueda2008tfs,liu2007ats,machida2006gpd}
and also appears in balanced BCS theory when studying a
normal-superfluid interface. We believe this is not a signature
of an FFLO phase in the gas, because for that the homogeneous
phase diagram should contain a so-called Lifshitz point where
the superfluid density becomes negative and it is energetically
favorable to form Cooper pairs with a nonzero center-of-mass
momentum. But we have checked that this theory at unitarity
does not contain such a Lifshitz point \cite{jildou2009lpp}.
The oscillations are also not related to the self-energy
contributions, however, the oscillations are enhanced by it as
the self-energies depend on the gap parameter. In the current
theory, no extra costs for gradients in the gap parameter are
included, which would clearly suppress these oscillations.

The problems with the Bogoliubov-de Gennes approach are
difficult, if not impossible to overcome within this framework.
We therefore believe that a different approach, using
Landau-Ginzburg theory, in which a gradient effects for the gap
parameter can be easily taken into account, works much better.
The position of the interface can then correctly be included by
choosing the correct self-energies. The oscillations will be
suppressed as a result of gradients contributions that
introduces an extra energy cost for rapid variations in the
gap. How this works in practice is discussed next.

\section{Landau-Ginzburg}

In the previous section we discussed the Bogoliubov-de Gennes
method to study beyond local-density behavior in the vicinity
of the superfluid-normal interface. We want to stress once more
that this method experiences problems in a system with a
first-order transition, that can not be readily resolved.
Therefore, to obtain a more detailed picture of the interface
we need a different approach. Because we want to describe a
first-order phase transition, the location of the transition,
and therefore the interface, is determined by the values of the
thermodynamic potential in its minima. To do this correctly we
need an appropriate thermodynamic potential that describes both
the normal and the superfluid phase. This thermodynamic
potential can then be extended with a gradient term to take the
energy cost of a varying gap parameter into account.

We arrive at our most accurate theory for the inhomogeneous
Fermi mixture in the unitarity limit by constructing an
approximation to the exact Landau-Ginzburg (grand-canonical)
energy functional $\Omega[\Delta(x);\mu,h]$ for the BCS gap
parameter $\Delta(x)$. Here, $\mu_+=\mu+h$ and $\mu_-=\mu-h$
are again the chemical potentials for the majority and minority
atoms respectively. The approach is very different from the
previous approach based on the Bogoliubov-de Gennes equations.
Although these equations exactly diagonalize the fermionic part
of the microscopic action, the BCS gap parameter is then only
obtained in the saddle-point approximation. As a result,
crucial fluctuation corrections are missing in the strongly
interacting limit. While self-energy corrections can still be
readily included in the Bogoliubov-de Gennes approach,
diagrammatic vertex corrections to the particle-particle
correlation function that also affect the gap equation are
neglected.

\subsection{Normal phase}

The Landau-Ginzburg approach is based on the central idea that
the relevant physics of the strongly interacting system is not
only captured by fermionic self-energy insertions, but that
also gradients of the order parameter are important. In this
context it is important to realize that there exists an exact
energy functional $\Omega[\Delta;\mu,h]$ that describes the
system exactly and that we thus want to approximate as best as
we can. As a first step towards this goal we can use mean-field
theory, because it is now rather well established that
mean-field theory gives a correct qualitative description of
the unitarity limit. Namely, at zero temperature both
experiments and several Monte Carlo calculations find a
first-order transition at a local critical imbalance of about
$P_c=0.4$, while for the balanced case both find a second-order
phase transition at about $T_c=0.15 T_F$. The second and
first-order transitions should then be connected by a
tricritical point as confirmed experimentally. Since the BCS
energy functional gives rise to the same qualitative behavior
it must have the same shape and temperature dependence as the
exact functional. We therefore start with BCS theory, after
which we include the dominant effects we are still missing. The
BCS energy functional is
\begin{align}\label{eq:BCSfreeEnergy}
    \Omega_{\text{BCS}}[\Delta; \mu, h] =&
    \sum_{\bm k}\left(\epsilon_{\bm k}-\mu-\hbar\omega_{\bm k}
                  +\frac{|\Delta|^2}{2\epsilon_{\bm k}}\right)\\ \nonumber
    &-k_{\text{B}}T\sum_{\bm k, \sigma}
      \log\left(1+\ee^{-\hbar\omega_{\bm k, \sigma}/k_{\text
      B}T}\right)~,
\end{align}
where $\epsilon_{\bm k} = \hbar^2\bm k^2/2m$, $m$ is the mass,
and the superfluid dispersion is $\hbar\omega_{\bm
k}=\sqrt{(\epsilon_{\bm k}-\mu)^2+|\Delta|^2}$. The second sum
is over the pseudo-spin $\sigma=\pm$ with $\hbar\omega_{\bm
k,\sigma} = \hbar\omega_{\bm k} - \sigma h$.

Since BCS theory incorrectly leads to a noninteracting normal
state, we first incorporate the fermionic self-energy effects,
such that we are able to correctly describe the strongly
interacting normal state. To incorporate self-energy effects,
we have to replace the chemical potentials by their
renormalized versions
$\mu_{\sigma}'=\mu_{\sigma}-\hbar\Sigma_{\sigma}$, where
$\hbar\Sigma_{\sigma}$ are the appropriate fermionic
self-energies. Inspired by Hartree-Fock theory we take in the
normal state the following Ansatz \cite{gubbels2006spt}
\begin{align}\label{eq:LDA_SelfEnergy}
    \mu_{\sigma}'=\mu_{\sigma} +
    \frac{3}{5}A\frac{(\mu_{-\sigma}')^{2}}{\mu_+'+\mu_-'}\;.
\end{align}
From the Monte Carlo result for a fully polarized gas
\cite{lobo2006nsp}, we know that minority particles start to
appear at zero temperature when $\mu_-=-0.6\mu_+$. This fixes
the prefactor in the Ansatz for the self-energy which is then
the same as in \eqref{eq:selfenergy_density}. This construction
leads to excellent agreement with the full Monte Carlo equation
of state. This is very nicely illustrated in Fig 2.~of
Ref.~\cite{jildou2009lpp} (dashed line) were the same
self-energy is used. As a result, we are thermodynamically
completely equivalent to Monte Carlo calculations in the normal
state at zero temperature. Moreover, we have also checked that
our construction leads to the correct Monte Carlo result for
the balanced gas at $T_c= 0.15 T_{\text{F}}$. Namely, our
construction also gives $\mu=0.5
\epsilon_{\text{F}}$ just as found in
Ref.~\cite{burovski2006ctt}. This agrees with our assumption
that the coefficient $A$ does not depend too much on
temperature in the normal state.

This form of the self-energy is closely related to the one in
\eqref{eq:selfenergy_density}, which can be seen by replacing
the densities with the ideal gas value. These two different
forms behave similar for the normal phase, but for the
superfluid phase it is clear that the self-energy in
\eqref{eq:LDA_SelfEnergy} has not the correct behavior. To
account for this we will introduce the necessary corrections in
the next subsection.

\subsection{Superfluid phase}

We now turn to the superfluid state. Since at zero temperature
a phase separation occurs between the normal state and a
balanced superfluid, the (diagonal) self-energy in the balanced
superfluid is also important for our purposes. Since both our
normal state construction and the use of BCS theory take
interaction effects into account in the superfluid state, there
are double counting problems. To correct for this, the strategy
is again too match the superfluid Monte Carlo result, so that
the coefficient of our extra superfluid self-energy correction
is uniquely fixed by the known value for the energy in the
superfluid minimum. In the superfluid phase it is much harder
to calculate self-energy corrections from first principles,
however, what can be shown analytically is that corrections to
the self-energy in the superfluid state can be expanded as a
series with even powers in the gap parameter. As a result, the
direct way to account for the superfluid self-energy effects is
by matching a second-order correction in the BCS gap parameter
\cite{bulgac2008ufs} to Monte Carlo. This can be accomplished
by adding $\varDelta\Sigma_{\sigma}=0.21
\Delta^2/\hbar\mu'$ to the self-energies. The proportionality
constant is fixed by the requirement that in the balanced case
the minimum of the energy functional equals
\begin{align}\label{eq:SF_energyfunc_minimum}
    \Omega_{\text{cr}}/V = -\frac{4\sqrt{2} \mu
    ^{5/2}m^{3/2}}{15 \pi ^2 \hbar^3(1+\beta)^{3/2}}\;,
\end{align}
with $\beta=-0.58$ and $V$ the volume. Also, the critical
imbalance for phase separation is now automatically
incorporated exactly into the theory at zero temperature.

\begin{figure}[t]
    \includegraphics[width=\columnwidth]{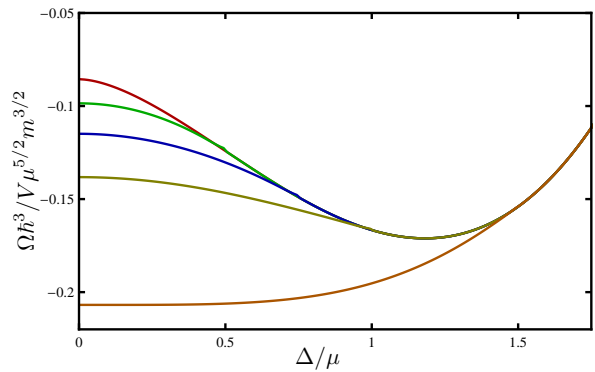}
    \caption{(Color online) The
    zero temperature energy functional as a function of the order
    parameter $\Delta$ with the self-energy of \eqref{eq:BdG_SF_SelfEnergy}
    for different values of $h$.}\label{fig:densityEnergyFunctionals}
\end{figure}

\begin{figure}[t]
    \includegraphics[width=\columnwidth]{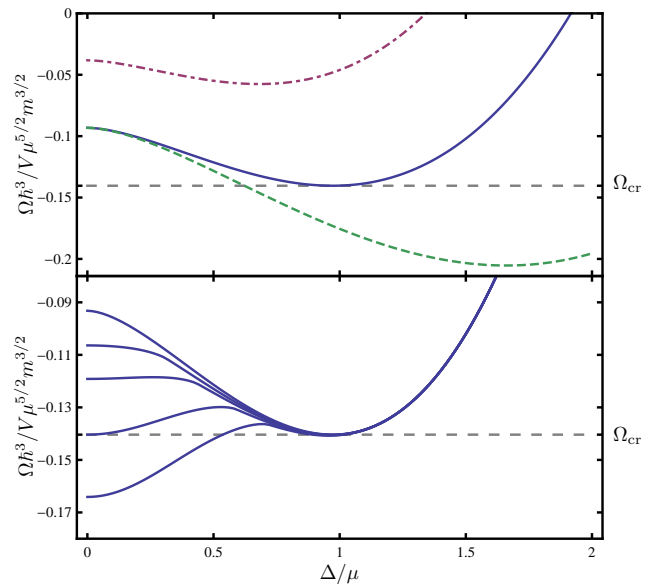}
    \caption{(Color online) The
    zero temperature energy functional as a function of the order
    parameter $\Delta$. The upper panel illustrates the balanced case,
    where the dash-dotted line is the usual BCS result, the dashed
    line incorporates only the normal state self-energy effects, and
    the solid line includes also the superfluid
    $\varDelta\Sigma_{\sigma}$ correction. In the lower panel the
    energy functional is shown for various values of the chemical
    potential difference $h$, with $h_{\text{cr}}=0.94\mu$ its
    critical value.}\label{fig:energyFunctionals}
\end{figure}

At this point our construction gives rise to one last problem,
namely that the terms in the thermodynamic potential that
describe the condensate of Cooper pairs, still depends on $h$
in the superfluid state through the renormalization $\mu'(\mu,
h)$. This is not correct, because at zero temperature the phase
separation occurs between a partially polarized normal state
and an unpolarized superfluid. To solve this problem, we have
chosen to exponentially suppress the $h$ dependence in the
superfluid state, which then finally leads to the correct shape
of the thermodynamic potential in all known limits. Technically
this is achieved by writing $\mu'$ in terms of $\mu$ and $h$
while exponentially suppressing the $h$ dependence through the
substitution $h\rightarrow h \exp{(-4 |\Delta|^2/\mu^2)}$. The
factor of 4 in the exponent of the suppression is somewhat
arbitrary, but it is large enough to make the $h$ dependence in
the ground state superfluid minimum vanish. Slight variations
in this factor do not affect our results. Note that the whole
problem is truly artificial, since if we would have calculated
the self-energy corrections in terms of the densities rather
than the chemical potentials, then all $h$-dependence would
have been automatically exponentially suppressed in the
superfluid state. This is what was done in the previous part
using \eqref{eq:BdG_SF_SelfEnergy} and is shown explicitly in
Fig.~\ref{fig:densityEnergyFunctionals}, where the
energy-functional is plotted for the density dependent
self-energy in \eqref{eq:selfenergy_density}. The
energy-functional we just constructed with the self-energy
given above \eqref{eq:LDA_SelfEnergy} is shown in
Fig.~\ref{fig:energyFunctionals} and has the same behavior as a
function of $h$ as in Fig.~\ref{fig:densityEnergyFunctionals}.
This thus proves that this suppression of the $h$ dependance
captures the correct and relevant physics. The reason that the
energy functional with the self-energy of
\eqref{eq:LDA_SelfEnergy} is preferred, is that it does not
directly depend on the densities. This makes it consistent with
the use of the grand-canonical ensemble and much easier to use.

Now our theory gives the correct equation of state in both the
superfluid and the normal phase and thus also the correct
critical polarization. Even the outcome for the universal
number $\zeta=\Delta/\mu$ of the balanced superfluid ground
state is reasonable. Here, we find $0.97$ while Monte Carlo
gives $1.07\pm0.15$ \cite{carlson2008spg,carlson2005atf}.
Moreover, our functional also provides a theoretical
description of the system in case the order parameter is not in
a minimum of the thermodynamic potential as illustrated in
Fig.~\ref{fig:energyFunctionals}. This is very important for
our purposes as it can be used to study also the
superfluid-normal interface.

\begin{figure}[t]
    \includegraphics[width=\columnwidth]{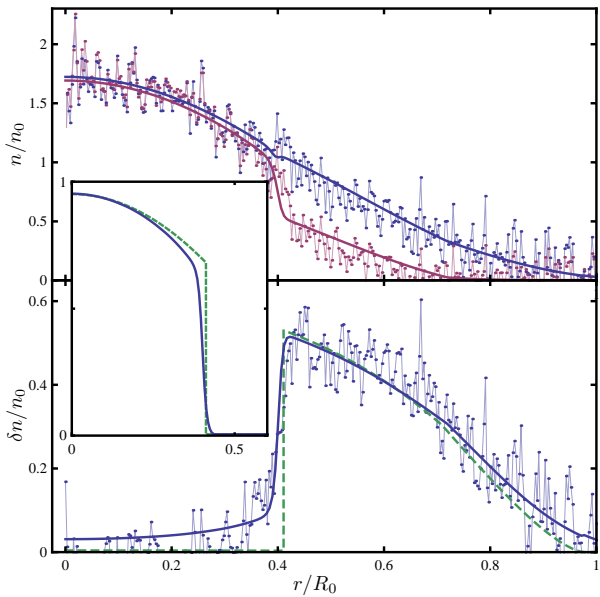}
    \caption{
    (Color online) The density profiles
    of a unitary mixture with polarization P=0.44 in a harmonic trap.
    The upper figure shows the majority and minority densities as a
    function of the position in the trap. The lower figure shows the
    density difference, where the theoretical curves show the results
    both within LDA (dashed line) and for our theory (solid line) that
    goes beyond this approximation and, therefore, allows for a
    substantial better agreement with experiment. The inset shows the
    BCS gap parameter $\Delta(r)$ both for LDA (dashed line) and our
    theory (solid line). The experimental data (dots) and
    scaling is from Shin {\it et al}.\ \cite{shin2008pdt}.
    }\label{fig:profiles}
\end{figure}

To go beyond LDA, we now take into account the gradient term in
the energy functional, resulting in
\begin{align}\begin{split}
    \Omega[\Delta;\mu,h] =& \Omega_{\text{BCS}}[\Delta;\mu',h']\\
       &+ \frac{1}{2} \int \dd \bm{x}~
            \gamma(\mu,h)|\nabla\Delta|^2\;,
\end{split}
\end{align}
where $\hbar \gamma(\mu,h) \sqrt{\mu/m}$ is a positive function
of the ratio $h/\mu$ only. This changes the discontinuous step
at the interface obtained within LDA, into a smooth transition.
A careful inspection of the interface in the data of Shin {\it
et al}.\ \cite{shin2008pdt}, cf.\ Fig.~\ref{fig:profiles}, also
reveals that the interface is not a sharp step. This is most
clear in the data of the density difference, since the noise in
the difference is much smaller than in the total density. This
has to do with the experimental procedure used, which
independently measures the total density and density
difference. As we have seen, a smooth transition arises also in
the self-consistent Bogoliubov-de Gennes equations. But these
lead then also to oscillations in the order parameter and the
densities, due to the proximity effect \cite{mcmillan1968tsn}.
This is not observed experimentally. Oscillations will also
occur in our Landau-Ginzburg approach if $\gamma(\mu,h)<0$.
However, we have checked both with the above theory as well as
with renormalization group calculations \cite{gubbels2008rgt}
that $\gamma(\mu,h)$ is positive. This agrees with the phase
diagram of the imbalanced Fermi mixture containing a
tricritical point and not a Lifshitz point in the unitarity
limit \cite{jildou2009lpp}.

We restrict ourselves here to a gradient term that is of second
order in $\Delta$ and also of second order in the gradients.
There are of course higher-order gradient terms that may
contribute quantitatively \cite{stoof1993tgl}, but the leading
order physics is captured in this way due to the absence of a
Lifshitz transition. One way to compute the coefficient
$\gamma(\mu,h)$ is to use the fact that in equilibrium this
coefficient can be exactly related to the superfluid stiffness,
and therefore the superfluid density $\rho_{\text{s}}$, by
$\gamma =\hbar^2\rho_{\text{s}}/4m^2|\langle\Delta\rangle|^2$.
At zero temperature it gives the simple result that
$\gamma(\mu,h)=
\sqrt{m/2\mu}/6\pi^2\hbar\zeta^2(1+\beta)^{3/2}$, with $\beta$
and $\zeta$ universal constants as defined earlier.

\section{Comparison with Experiments}

\begin{figure}
    \includegraphics[width=\columnwidth]{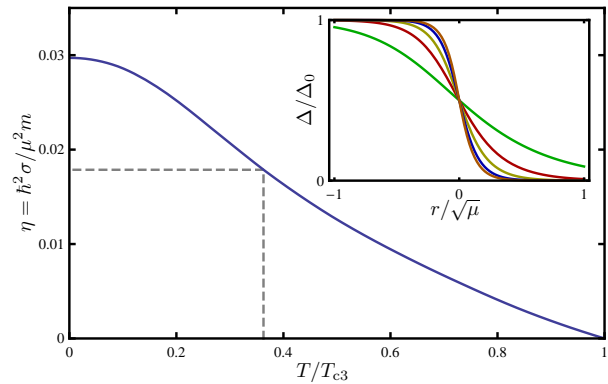}
    \caption{
    (Color online) The surface tension as a function of the temperature, computed
    in the homogeneous case at unitarity. The temperature
    is scaled by the temperature of the tricritical point. The dashed line shows
    the value used to compare with experiments. The inset shows the gap around the
    interface for several temperatures ($0.9$, $0.7$, $0.5$, $0.25$ and $0.01$ $T_{\text{c}3}$).
    }\label{fig:LG_SurfaceTension}
\end{figure}

Up to now we have focussed on the zero temperature limit.
However, our arguments are also valid for nonzero temperatures
$T \ll T_{F+}$, with $T_{\text{F+}}$ the Fermi temperature of
the majority species. Here we are allowed to neglect the
temperature dependence of the self-energies and the superfluid
density. The data of Shin {\it et al}. of interest to us is
taken at a temperature of $0.03T_{\text{F+}}$. This is about a
third of the temperature at the tricritical point
$T_{\text{c3}}$. A nonzero temperature significantly affects
the surface tension and increases the width of the interface,
because it lowers the energy barrier between the normal and
superfluid phases. Indeed at the tricritical point this
barrier, and thus the surface tension, exactly vanishes. We
therefore also perform calculations at $0.3 T_{\text{c3}}$. As
mentioned, the leading-order temperature effects are
incorporated in the BCS energy functional of
\eqref{eq:BCSfreeEnergy}.

\subsection{Surface Tension}

The fact that we are able to study the superfluid normal
interface beyond LDA, makes it possible for us to also
determine the surface tension. The surface tension is given by
the (grand-canonical) energy difference between a
one-dimensional LDA result with a discontinuous step in
$\Delta(z)$ and our Landau-Ginzburg theory with a smooth
profile for the order parameter $\Delta(z)$. To write the
surface tension $\sigma$ in a dimensionless form, we use
$\sigma = \eta (m/\hbar^2)\mu^2$, with $\eta$ a dimensionless
number. In this form it was previously found that for the Rice
experiment $\eta=0.6$ \cite{hulet2006pps}. This was extracted
from the large deformations of the superfluid core observed in
that experiment. The experiment of Shin {\it et al}.\
\cite{shin2008pdt} does not show any deformation, which puts an
upper bound on $\eta$ of about $0.1$
\cite{haque2007tfc,baur2009tns}.

The size of the interface is rather small compared to the size
of the whole trap. This makes it possible to compute the
surface tension of the interface in a homogeneous system rather
than in the whole trap. In Fig.~\ref{fig:LG_SurfaceTension} the
surface tension of this model is plotted as a function of the
temperature. At the tricritical point the surface tension
vanishes and at zero temperature it is about $\eta=0.03$. For a
more realistic temperature of about $0.3 T_{\text{c3}}$ we find
$\eta=0.02$ which is significantly smaller than the critical
surface tension that would cause deformation. This is thus in
agreement with the MIT experiment.

We now give a more detailed discussion of our analysis of the
density profiles observed by Shin {\it et al}. In experiments
the cloud is trapped in an anisotropic harmonic potential,
which is in the axial direction less steep than in the radial
direction. However, since the gas cloud shows no deformations
in this case we can in a good approximation take the trap to be
spherically symmetric. The order parameter then depends only on
the radius and the total Landau-Ginzburg energy is given by
integrating the Landau-Ginzburg energy density over the trap
volume. To account for the trap potential in the energy
functional we let the chemical potential depend on the radius,
such that we have $\mu_{\sigma}(r) =
\mu_{\sigma} - V(r)$, with $V(r)$ the effectively isotropic
harmonic potential.

To find the order parameter as a function of the radius we have
to minimize the energy functional with respect to the order
parameter, thus $\delta\Omega[\Delta;\mu,h]/\delta\Delta(r)=0$.
This gives a second-order differential equation for
$\Delta(r)$. Solving this Euler-Lagrange equation, with the
proper boundary conditions in the center of the trap, gives a
profile for $\Delta$ as is shown in the inset of
Fig.~\ref{fig:profiles}. This profile of the order parameter is
much smoother than the discontinuous step one obtains within
LDA that is also shown in Fig.~\ref{fig:profiles}. Besides
this, there are two more aspects that deserve some attention.
First, we notice that the value of the gap at the original LDA
interface is decreased by almost a factor of three and, second,
the gap penetrates into the area originally seen as the normal
phase. This behavior makes the gap for a small region smaller
than $h'$, giving locally rise to a gapless superfluid, which
implies a stabilization of the Sarma phase.

Before discussing this particular physics, we focus first on
the density difference. To obtain the density profiles within
our theory, the thermodynamic relation
$n_{\sigma}(r)=\partial\Omega/\partial\mu_{\sigma}(r)$ is used,
where $n_{\sigma}$ is the density of particles in state
$\sigma$ and $\mu_{\sigma}(r)$ the associated local chemical
potential. It is important that, because of the self-energy
effects, we cannot use the standard BCS formulas for the
density, but really have to differentiate the energy
functional. In BCS theory this would of course be equivalent.
Given the density profiles, the comparison between theory and
experiment can be made and is ultimately shown in
Fig.~\ref{fig:profiles}. Overall the agreement is very good.
Theoretically the interface appears to be somewhat sharper than
observed. This can be due to higher-order gradient terms, that
are neglected in the calculation and that would give an
additional energy penalty for a spatial variation of the order
parameter. There are however experimental effects that could
make the interface appear broader, for instance, the spatial
resolution of the tomographic reconstruction or the accuracy of
the elliptical averaging \cite{priv:ket}.

The Landau-Ginzburg approach presented here, shows some new
features compared to LDA. One interesting feature is the kink,
that is clearly visible in the majority density profile shown
in Fig.~\ref{fig:profiles}. Notice that this kink appears
\emph{before} the original (LDA) phase transition from the
superfluid to the normal phase. This kink signals a crossover
to a new exotic phase, namely the gapless Sarma phase. Note
that at zero temperature it becomes a true quantum phase
transition. At the crossover, the order parameter becomes
smaller than the renormalized chemical potential difference
$h'$ and the unitarity limited attraction is no longer able to
fully overcome the frustration induced by the imbalance. As a
result the gas becomes a polarized superfluid. Because the gap
$\Delta$ is smaller than $h'$ this corresponds to a gapless
superconductor. In a homogeneous situation this can, far below
the tricritical temperature, never be a stable state. However,
because of the inhomogeneity induced by the confinement of the
gas, the gap is at the interface forced to move away from the
local minimum of the energy functional and ultimately becomes
smaller than $h'$. The Sarma state is now locally stabilized
even at these low temperatures. Notice that this is a feature
of a smooth behavior of the gap and that the presence of the
Sarma phase does not depend on the quantitative details of the
energy functional $\Omega[\Delta ;\mu, h]$.

\section{Conclusions}

In this paper we first studied the Bogoliubov-de Gennes method
to go beyond LDA. We showed that the addition of a self-energy
gives a model that reproduces the known Monte-Carlo results of
the homogeneous system. However, we argued that this
Bogoliubov-de Gennes approach suffers from some fundamental
difficulties and that an approach using Landau-Ginzburg theory
gives more accurate results. We used results from Monte-Carlo
calculations to construct an approximation to the exact
Landau-Ginzburg energy functional that describes the
experimental data without any free parameters. We considered
beyond LDA effects on imbalanced Fermi mixtures in the
unitarity limit and showed that this results in a much better
agreement with experiments than LDA. The interface details will
depend on both the polarization and temperature, but there is
not sufficient experimental data available for such a
systematic study. Moreover, we found that exotic physics is
occurring in the superfluid-normal interface. We also showed
that an experimental signature of the gapless Sarma phase is a
kink in the majority density profile. The temperature plays an
important role here, since a lower temperature will lead to a
more visible kink but also to a sharper interface. Presumably a
compromise will have to be found in this respect. We hope that
our work will stimulate more experimental work in this
direction.

\begin{acknowledgments}
We thank Achilleas Lazarides, Randy Hulet, Wolfgang Ketterle,
and Yong-il Shin for stimulating discussions and for kindly
providing us with their data. This work is supported by the
Stichting voor Fundamenteel Onderzoek der Materie (FOM) and the
Nederlandse Organisatie voor Wetenschaplijk Onderzoek (NWO).
\end{acknowledgments}

\end{document}